\documentclass[9pt,twocolumn,twoside]{osajnl}

\journal{ao} 

\setboolean{shortarticle}{false}

\title{Curved detectors for astronomical applications:\\characterization results on different samples}

\author[1,*]{Simona Lombardo}
\author[1]{Thibault Behaghel}
\author[2]{Bertrand Chambion}
\author[2]{St\'ephane Caplet}
\author[3]{Wilfried Jahn}
\author[1]{Emmanuel Hugot}
\author[1]{Eduard Muslimov}
\author[1]{Melanie Roulet}
\author[1]{Marc Ferrari}
\author[2]{Christophe Gaschet}
\author[2]{David Henry}

\affil[1]{Aix Marseille Univ, CNRS, CNES, LAM, Marseille, France}
\affil[3]{Division of aerospace engineering, Caltech, Pasadena, CA 91125, USA}
\affil[2]{Univ. Grenoble Alpes, CEA, LETI, MINATEC campus, F38054 Grenoble, France}

\affil[*]{Corresponding author: simona.lombardo@lam.fr}

\begin{abstract}
 Due to the increasing dimension, complexity and cost of the future astronomical surveys, new technologies enabling more compact and simpler systems are required. The development of curved detectors allows to enhance the performances of the optical system used (telescope or astronomical instrument), while keeping the system more compact. 
 We describe here a set of five curved CMOS detectors developed within a collaboration between CEA-LETI and CNRS-LAM.
These fully-functional detectors 20\,Mpix (CMOSIS CMV20000) have been curved to different radii of curvature and spherical shapes (both convex and concave) over a size of 24x32\,mm$^2$. Before being able to use them for astronomical observations, we assess the impact of the curving process on their performances.
We perform a full electro-optical characterization of the curved detectors, by measuring the gain, the full well capacity, the dynamic-range and the noise properties, such as dark current, readout noise, pixel-relative-non-uniformity. 
   We repeat the same process for the flat version of the same CMOS sensor, as a reference for comparison. 
We find no significant difference among most of the characterization values of the curved and flat samples. We obtain values of readout noise of 10e$^-$ for the curved samples compared to the 11e$^-$ of the flat sample, which provides slightly larger dynamic ranges for the curved detectors.
   Additionally we measure consistently smaller values of dark current compared to the flat CMOS sensor.
   The curving process for the prototypes shown in this paper does not significantly impact the performances of the detectors. These results represent the first step towards their astronomical implementation.
\end{abstract}

\setboolean{displaycopyright}{true}

\begin{document}

\maketitle

\section{Introduction}
\label{sec:intro} 
In the recent years the need for technological progress in astronomy has been growing faster and faster, leading to more demanding surveys in terms of mechanical and optical complexities, such as the Extremely Large Telescope (ELT), the Thirty Meter Telescope (TMT), the Giant Magellan Telescope (GMT), The Large UV/Optical/IR Surveyor (LUVOIR), etc \citep{tmt2013,gmt2014,luvoir_2018}.
Thus, the necessity of developing innovative systems that allow to reduce complexities, dimensions and costs without impacting the performances has become imperative.
\begin{figure*}[ht]
 \begin{center}
  \includegraphics[width=0.90\textwidth]{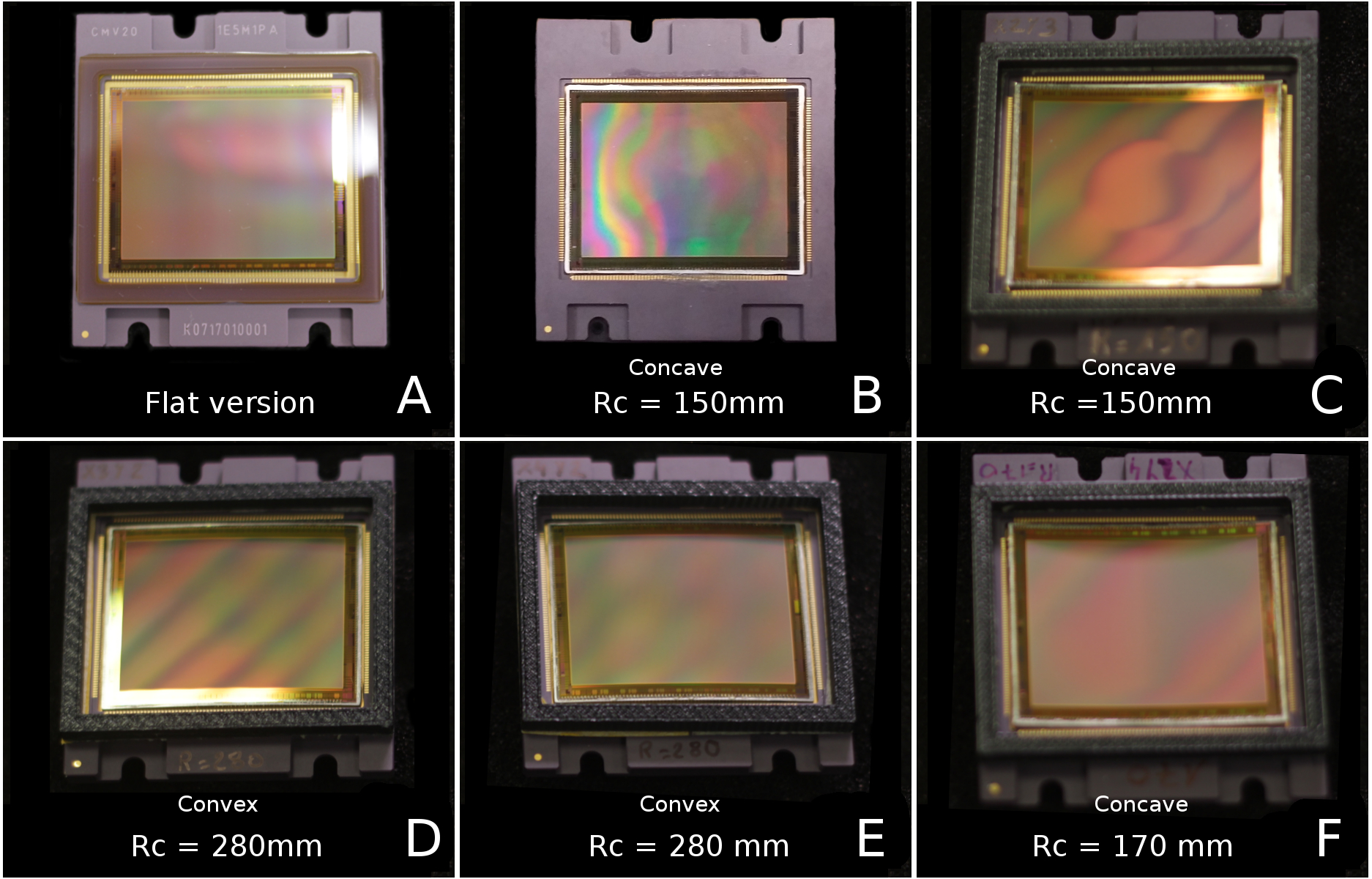}
  \caption{Six CMV20000 CMOS image sensors from CMOSIS. Sample A is the flat off-the-shelf component and all the others have been curved with a spherical shape to the radius of curvature listed in Table~\ref{tab: samples}.}
  \label{four_prototypes}
 \end{center}
\end{figure*}
The research and development have been focusing on many different aspects, from the 3D printing of lightweight structure of astronomical mirrors \citep{atkins2017} to freeform surfaces.
The study of curved detectors also belongs to this framework \citep[curved CCDs\footnote{\url{http://www.andanta.de/pdf/andanta_ccd_curved_overview_en.pdf}} and CMOS][]{guenter2017,iwert_2012,dumas_2012,tekaya_2013,Itonaga_2014,gregory_2015}.
This technology has been gathering increasing attention as the fields of application are numerous, from low-cost commercial to high impact scientific systems, to mass-market and on board cameras \citep{swain_2004}, defense and security \citep{tekaya_2014}. 

In astronomy, the possibility of having a sensor for a curved focal plane allows to explore a larger parameter space for the optical design and to find solutions with improved performance on several criteria such as: homogeneity and quality of the Point Spread Function (PSF) in the field, general distortion of the image and chromatic aberrations. 
Many optical systems, especially those with wide field of views, generate curved focal planes that require additional optical elements (field-flattener) to project the image on flat detectors.
In the astronomical domain we can for instance cite Kepler \citep{kepler2010} and the Zwicky Transient Facility \citep[ZTF,][]{ztf2016}. 
Therefore, in order to obtain the correct image, designers have to compromise the throughput and the performance of their systems.
However this is no longer necessary when using a curved detector.
The systems become consequently more compact and simpler, delivering at the same time better performances (e.g. increased resolution).
This is particularly advantageous for space missions.

As several prototypes of curved detectors have been already produced \citep{guenter2017,iwert_2012,dumas_2012,tekaya_2013,Itonaga_2014}, their integration in astronomy is about to become a reality. 
There are ongoing plans to use this technology in future instrumentation. 
The proposed satellite mission MESSIER \citep{valls-gabaud2017}, for example, would greatly benefit from using curved detectors.  
MESSIER aims at measuring surface brightness levels as low as 35 mag arcsec$^{-2}$ in the optical (350-1000 nm) and 38 mag arcsec$^{-2}$ in the UV (200 nm).
For its design, any refractive surface must be excluded, as they would generate Cherenkov emission due to the relativistic particles (hence, no field flattening optics are allowed).
Additionally, as the goal of the space-based telescope is to observe the ultra-low surface brightness universe, the instrumental PSF must be as compact as possible, while guaranteeing a wide field of view.

\cite{eduard2017} have proposed a demonstrator for this satellite, with a small telescope (35\,cm diameter of primary mirror).
As the ground-based pathfinder must be as close as possible to the space-based version, it has been designed as a fully reflective Schmidt telescope with a convex focal plane and a radius of curvature of 800\,mm (equipped with a curved CCD).
The selected observing mode is drift scan, which requires a distortion free PSF in the scanning direction across the field of view of $1.6^o\times2.6^o$. 
By using curved detectors the PSF is considerably less distorted in the edge of the field of view, with respect to the design case with field flattening optics and flat detector.

Curved detectors have been proposed also for BlueMuse (Richard et al., in prep), an Integral Field Spectrograph (IFS) which will complement the science done with the current Multi-Unit Spectrosopic Explorer \citep{muse2016}, while exploring bluer wavelengths.
By allowing the focal plane to be curved, the optics of the planned BlueMuse become smaller and the overall optical design slimmer.
In this paper, we describe a set of five curved CMOS detectors developed in the frame of a collaboration between CNRS-LAM and CEA-LETI (Section~\ref{sec:sections}).
This fully-functional front-side illuminated detectors of 20\,Mpix (CMOSIS, CMV20000\footnote{\url{http://www.cmosis.com/products/product_detail/cmv20000}}) have been curved to different radii and with different shapes over a size of 24x32\,mm$^2$ (full-frame sensor sensitive to the visible light). 
After the curving process, these chips were repackaged in the same packaging as the original one before curving, in such a way that the final product is a “plug-and-play” component.

To allow the full exploitation of curved detectors for the astronomical community, the first step is to test the impact of the curving process on their performances.
Hence, after having calibrated their internal temperature sensors (Section~\ref{sec:temp_cal_intro}), we present here the methodology adopted for their characterization (Section~\ref{sec:met1}) in terms of noise properties -- such as readout noise, dark current and pixel-relative-non-uniformity -- and gain.
In Section~\ref{sec:results}, we present all the results obtained and compare these to the results from the characterization (performed with the same methodology) of a flat version of the same detector. 
We conclude in Section~\ref{sec:conclusion}.

\section{Process for curving CMOS}
\label{sec:sections}

Several curved detector concepts are being prototyped at CEA-LETI in collaboration with CNRS-LAM.
Some of these have already proved the improvements achievable in terms of compactness and performances of the related optical designs \citep{cea_2018}. 
In this section we provide some detail regarding this successful curving process. 

The initial flat sensor consists of a silicon die glued on a ceramic package, where the electrical connections are provided by wire bonding from the die to the package surface. 
Additionally, a glass window protects the sensor surface from mechanical or environmental solicitations. 
The curving process of these sensors consists of two steps: firstly the sensors are thinned with a grinding equipment to increase their mechanical flexibility, then they are glued onto a curved substrate. 
The required shape of the CMOS is, hence, given by the shape of the substrate.
The sensors are then wire bonded in a way that they keep the packaging identical to the original one before curving.
The final product is, therefore, a component ready to be used or tested.

In this paper we show the electro-optical characterization results of five of our prototypes.
These chips (Figure~\ref{four_prototypes}) are CMV20000 global shutter CMOS image sensors from CMOSIS, with 5120$\times$3840 pixels of 6.4\,$\mu$m size. 
They have been curved with spherical shapes at different radii of curvature as listed in Table~\ref{tab: samples}.
Having now 5 curved samples, we can have a statistical analysis and draw more robust conclusions.
The analysis in this paper can only be used as a statistical way (on all the samples) of testing the properties of the sensors after the curving process, as their properties before curving and thinning are unknown.
In order to test the flat sensors, before any modification is applied, they would have to be packaged and wire bonded. This, however, would pose a high risk of damaging the sensors themselves when the package and wires are removed to proceed with the thinning and curving process. Such risk was avoided for this study.
\begin{table}[h]
\begin{center}
\caption{List of CMOS samples tested in this paper, with their shape and radius of curvature ($R_\mathrm{c}$).}
\begin{tabular}{ccc}
\hline\\[-1.8ex]
   Sample name & Shape & $R_\mathrm{c}$ (mm)\\
\hline\\[-1.8ex]
A  &   flat & $\infty$ \\[0.25ex] 
B  &   concave & 150 \\ [0.25ex]
C  & concave & 150  \\ [0.25ex]
D  & convex & 280\\ [0.25ex]
E  & convex & 280\\ [0.25ex]
F & concave & 170 \\ [0.25ex]
\hline\\[-1.1ex]
\end{tabular}
\label{tab: samples}
\end{center}
\end{table}

\section{Calibration of internal temperature sensor}
\label{sec:temp_cal_intro}

Some of the quantities that characterize a detector are highly depending on the temperature of the die itself (e.g. the dark current), thus a crucial step is to determine its temperature while the measurement is performed.
CMV20000 detectors have internal temperature sensors that can be used to monitor the chip temperature.
Two registers -- register 101  and 102 -- are readout from the chip by using the CMV20000 Evaluation Kit (Figure~\ref{termocoupl}a). 
The values of these registers are related to the temperature of the chip as follows:
\begin{equation}
T = l\times(256\times r_{102}+r_{101}) + q,
\label{eq:temp_cal}
\end{equation}
where $r_{102}$ and $r_{101}$ are the register values.
As $l$ and $q$ are different for each CMOS sensor, we must provide an independent temperature measurement of the die and relate that to the register values as in 
Equation~\ref{eq:temp_cal}.

\subsection{Internal temperature sensor calibration: methodology and set-up}
We used a set of four thermocouples type K connected to a Quad MAX31856 board, that converts the output of the thermocouples in temperature values.
We powered it and read it out with an Arduino MKR1000.

\begin{figure}[!h]
 \begin{center}
  \includegraphics[width=0.5\textwidth]{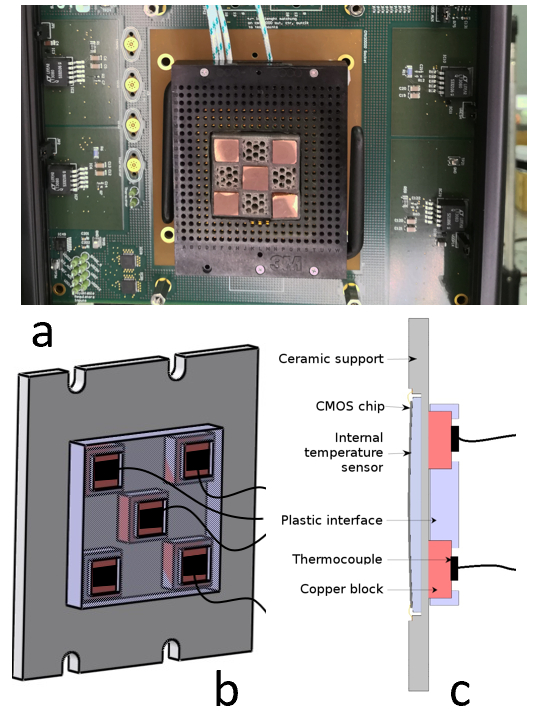}
  \caption{a: temperature calibration system installed inside the CMV20000 Evaluation Kit. The green/white cables are the thermocouples glued at the back of the copper blocks.  b: schematic view from the back of the temperature calibration system where the thermocouples are glued. c: schematic view from the side of the temperature calibration system where the position of the CMOS and of the ceramic support is shown from the left.}
 \label{termocoupl}
 \end{center}
\end{figure}

The thermocouples were glued to the back of four distinct copper blocks as in Figure~\ref{termocoupl}b.
These blocks have dimensions of 8\,mm$\times$8\,mm and a thickness of 3.2\,mm.
They sample the temperature at the back of the ceramic support of the CMV20000 chip in a region corresponding to its sensitive area (Figure~\ref{termocoupl}c).
For the flat CMOS sensor this ceramic support is located immediately below the chip.
The curved CMOS, however, additionally have the substrate used in the curving process (Section~\ref{sec:sections}), which is between the chip and the ceramic support.
For the curved samples C, D, E and F this substrate is made of invar, that facilitates heat dissipation, for sample B (the first prototype made) it is made of plastic. 
A difference in the temperature calibration of the samples is therefore expected. 

\begin{figure}[ht]
 \begin{center}
  \includegraphics[width=0.4\textwidth]{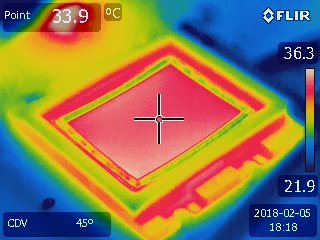}
  \caption{Image from the IR camera (FLIR I60BX) used to sample the temperature of the front surface of the CMOS sensors.}
  \label{IR_image}
 \end{center}
\end{figure}

Finally, a 3D-printed plastic structure holds the copper blocks together (Figure~\ref{termocoupl}).
The thermal link between the back of the ceramic support and the copper blocks was enhanced by applying some thermal grease at the top of the blocks.
By using this temperature calibration system -- the four thermocouples with all the mechanical supports and the electronics -- we have four different measurements of temperature from different areas of the back side of the CMOS.
In order to have an independent temperature probe of the front surface of the detectors, we also used an IR camera (FLIR I60BX, Figure~\ref{IR_image}).

The internal temperature sensor of each detector was calibrated singularly, by mounting the temperature calibration system at its back.
After powering the detector on, we readout at the same time: $r_{102}$ and $r_{101}$, the four thermocouples and we acquired an image with the IR camera pointing at the front of the sensor.
This image provides the temperature of the center of the CMOS sensor. 
We, then, let the sensor warm up (due to current flowing through it) and we repeated these steps again until it reached a stable temperature.

\subsection{Internal temperature sensor calibration: results}
\label{sec:cal_temp_res}

The temperature calibration system, provides a way to measure the change in temperature between two different measurements, but does not provide an estimate for the error on the absolute temperature value.
To test the accuracy of the thermocouples, we compared the temperature measured on average by the four thermocouples with the ones given by a mercury thermometer and a 
KIMO data logger (KISTOCK model KH210).
We readout the temperature probes at a refrigerator temperature ($\sim6^o$C) and we found that they all matched within the respective errors.
By repeating the test at room temperature ($\sim27^o$C) the result did not change.
We can, therefore, conclude that the absolute value of temperature provided by the thermocouples is accurate within $\pm0.5^o$C (the error here was the largest temperature mismatch found).

For each CMOS sensor described in this paper, we produced linear fits -- as in Equation~\ref{eq:temp_cal} -- by relating the values of register $r_{102}$ and $r_{101}$ to the temperature of the chip measured either with the IR camera or with the thermocouples value, averaged over all four of them.
For the flat sensor the presence of the protective window in front of the chip prevented us from acquiring the IR camera images. In this case only the thermocouples measurements were performed.

\begin{table*}[h]
\begin{center}
\caption{Values from the fit on the internal temperature sensor calibration for each of the CMOSIS sample analyzed in this paper. The second and third columns are from the fit from the average of the four values of the thermocouples and the last two columns are using only the thermocouple at the center of the system. The fourth and fifth columns are from the IR camera values.}
\begin{tabular}{lcccccc}
\hline\\[-1.8ex]
   & $l_{\mathrm{ter}}$  & $q_{\mathrm{ter}}$ & $l_{\mathrm{IR}}$ & $q_{\mathrm{IR}}$ & $l_{\mathrm{Cter}}$  & $q_{\mathrm{Cter}}$\\
\hline\\[-1.8ex]
A  &   0.32$\pm$0.01  & -357.3$\pm$13.5 & -- & -- & 0.330$\pm$0.008 & -371.4$\pm$10.6 \\[0.25ex] 
B  &   0.32$\pm$0.01 & -297.1$\pm$11.1 & 0.34$\pm$0.02 & -311.6$\pm$16.6 & 0.342$\pm$0.008 & -313.2$\pm$7.9 \\ [0.25ex]
C  & 0.321$\pm$0.003 & -293.9$\pm$3.4 & 0.33$\pm$0.01 & -298.9$\pm$9.3 & 0.338$\pm$0.006 & -310.1$\pm$5.9 \\ [0.25ex]
D  & 0.295$\pm$0.004 & -276.4$\pm$4.7 & 0.33$\pm$0.01 & -314.5$\pm$13.1 & 0.311$\pm$0.002 & -292.3$\pm$1.9 \\ [0.25ex] 
E  & 0.309$\pm$0.004 & -292.7$\pm$3.9 & 0.307$\pm$0.003 & -290.3$\pm$3.6 & 0.324$\pm$0.005 & -308.2$\pm$4.7 \\ [0.25ex]
F & 0.333$\pm$0.006 & -321.6$\pm$6.5 & 0.31$\pm$0.03 & -300.2$\pm$28.2 & 0.345$\pm$0.006 & -332.8$\pm$6.8 \\ [0.25ex]
\hline\\[-1.1ex]
\end{tabular}
\label{tab: values_temp}
\end{center}
\end{table*}
The results of the fits are presented in Table~\ref{tab: values_temp}.
The values of temperature obtained by substituting the IR camera parameters ($l_{\mathrm{IR}}$ and $q_{\mathrm{IR}}$) are always larger than what we obtain by substituting the thermocouples parameters ($l_{\mathrm{ter}}$ and $q_{\mathrm{ter}}$).
The last two columns of Table~\ref{tab: values_temp} show the parameters of the fit to the temperature measured by the thermocouple located at the center of the detector. 
Those values provide temperatures closer to the IR camera estimations, for most curved CMOS.

The difference between the value averaged over all four thermocouples and the value obtained by the central thermocouple can be explained by the presence of a gradient of temperature across the chip and its center being on average $\sim1^o$C hotter than the edges. 
As both the IR camera and the central thermocouple sample the temperature at the center of the CMOS sensor, their agreement shows that the temperature calibration system provides an unbiased measurement. 
However we still find a non negligible discrepancy for sample B.
Even when considering its central thermocouple values, we have $\sim2^o$C constant offset between these and the IR camera measurements.
This might be due to the presence of the plastic substrate between the CMOS chip and the ceramic support. 
Such substrate would prevent the dissipation of heat, generating in this way an offset in temperature.

As in this paper we are interested on characterizing the full sensitive surface of the detectors and not just its center, we used the temperature calibration results obtained from the average of all four thermocouples and kept the IR results as redundant test.
Once a reliable calibration for the internal temperature sensors was established, we used them as temperature monitor for each measurement performed during the  CMOS sensors characterization.

\section{Characterization of CMOS: tested quantities and methodology adopted}
\label{sec:met1}

The general aim of this paper is to characterize the CMOS sensors and evaluate the impact of the curving process on their performances. 
The measured characterization quantities include 6 main criteria \citep{janesick_2007,howell_2006}: gain of the detector, dark current, readout noise (RON), pixel response non-uniformity (PRNU), dynamic range (DR) and full well capacity (FW).
These features are briefly described in the following Section.

\subsection{Measured quantities for characterization}
\label{sec:meas_quant}
The dark current is due to the thermal agitation of the electrons within the semiconductor and it is extremely sensitive to the temperature of the detector itself. 
In Section~\ref{sec:temp_cal_intro} we established our own temperature calibration system which allowed us to calibrate the internal temperature sensor of the CMV20000 chip, used as temperature monitor during the characterization measurements. 

The dark current is a source of "unwanted" signal that puts stringent limits to the performances of the detector and it has to be carefully characterized and subtracted to any image, to improve its quality.
As the averaged number of counts on the detector grows linearly as a function of exposure time, the dark current is measured by reading out the detector at different exposure times in complete darkness conditions and a fit to these measurements is performed.
By computing the mean signal of a dark frame, readout after 0\,s of exposure time, we also obtain the bias level -- a positive offset due to the constant voltage applied by the electronic.

The readout noise (RON) is due to the scatter generated by the non perfectly reproducible conversion between analog to digital number, and the random fluctuation in the output signal introduced by the readout electronic.
The RON is obtained from the temporal noise, which in turn estimates the change of 
the output value of the pixels from contiguous exposures to a constant illumination level (valid also for dark frames).
As the temporal noise, $\sigma_{\mathrm{temp}}$ is composed of the RON and of the shot noise (due to the dark current or to the exposure to light) its value, when obtained from a set of dark frames at 0\,s exposure time, is equivalent to the RON itself.

The gain of a detector determines how many charges collected in each pixel are assigned to a digital number in the image.
The gain and the square of the temporal noise are in a linear relation as in the following Equation:
\begin{equation}
\sigma_{\mathrm{temp}}^2 = \mathrm{const} + k(S_{\mathrm{mean}}-S_{\mathrm{offset}}),
\label{eq:temp_noise}
\end{equation}
where $k$ is the gain in units of $\mathrm{DN/e}^{-}$ and $S_{\mathrm{mean}}-S_{\mathrm{offset}}$ are the mean signal of the frame and the bias level (mentioned before) respectively.
Here we assumed that the sensor is exposed to a uniform -- across the detector surface -- and stable illumination. 

For high illumination level (or longer exposure times), the PRNU starts becoming a dominant source of noise.
The pixels of a sensor have slightly different responses to incoming light and this effect generates the PRNU noise, which is directly proportional to the number of electrons detected, $N_e$, as in: $N_e = f_{\mathrm{PRNU}}\sigma_{\mathrm{PRNU}}$. The proportionality factor is called PRNU factor, $f_{\mathrm{PRNU}}$, and $\sigma_{\mathrm{PRNU}}$ is the PNRU noise.

In a frame, the total noise can be written as \citep{howell_2006}:
\begin{equation}
\sigma_{\mathrm{tot}} = \sqrt{\sigma_{\mathrm{e}}^2+\sigma_{\mathrm{RON}}^2+\sigma_{\mathrm{PRNU}}^2}
\label{eq:tot_noise}
\end{equation}
where $\sigma_{\mathrm{e}}$ is the photon noise, $\sigma_{\mathrm{RON}}$ is the readout noise, and $\sigma_{\mathrm{PRNU}}$ is the PRNU noise.
It should be pointed out that Equation~\ref{eq:tot_noise} only deals with RON to first order \citep[as indicated in][]{janesick_2007}.
By considering that the PRNU noise has a spatial dependence but not a temporal one, the subtraction of two frames, obtained at the same exposure time and the same uniform level of illumination, provides a new frame that contains only the RON and the photon noise.
From this the $\sigma_{\mathrm{e}}$ is obtained and by substituting it in Equation~\ref{eq:tot_noise}, $\sigma_{\mathrm{PRNU}}$ and $f_{\mathrm{PRNU}}$ are measured.

Finally the dynamic range -- the capability of the detector to be sensitive at high and low signal levels at the same time -- and the full well capacity -- the amount of charges a pixel can hold before saturating -- are defined as follows:
\begin{equation}
\mathrm{DR} = 20log(S_{\mathrm{max}}/\mathrm{RON}), \; \mathrm{FW} = S_{\mathrm{max}}-S_{\mathrm{offset}},
\label{eq:fwc}
\end{equation}
where $S_{\mathrm{max}}$ is the saturation limit, RON is the readout noise and $S_{\mathrm{offset}}$ is the bias level.

\subsection{Data acquisition}
\label{sec:data_acq}
A set of measurements was performed for each of the CMOS die listed in Section~\ref{sec:sections}.
The exposure time used for these tests varied between 0.0002\,s and the values at which saturation of the detector was reached, for exposures to uniform light -- also called flat fields -- and up to 0.96\,s for exposures in complete darkness -- or dark exposures. 

The measurements were made by acquiring frames with shorter and longer exposure times in mixed order. 
The alternation of these reduces (if not eliminates altogether) the impact of systematic effects due to light level drifts or small temperature fluctuations ($<0.1^{o}$C).
Thirty frames were acquired for each exposure time, with camera gain set to 1.
For each of the 30-block frames, the values of the internal temperature sensor were readout (Section~\ref{sec:temp_cal_intro}).

The setup used for the flat field frames included an integrating sphere, illuminated by a tungsten bulb located inside another smaller integrating sphere.
The integrating sphere was placed at a distance of 1.0 m from the sensor to achieve a uniform illumination of its surface.
Care was taken to reduce the scattered light, by using light baffles along the path from the integrating sphere to the detector housing. 

The measurements were performed at a room temperature of 21.0$\pm1.0^o$C, where the 
error is considered over the acquisition time for all the samples, $\sim5$\,days.
The variation during the full testing of a single detector was $\pm0.1^o$C.
The temperature of the CMOS chips was monitored by the internal temperature sensor, calibrated as explained in Section~\ref{sec:temp_cal_intro}.
Table~\ref{tab: measured_temp} shows the temperature measured in average during the full characterization of a single detector and the  errors in these estimates are the standard deviations from the average temperature measured.
For sample A, C, E and F the temperature was $\sim35.0^o$C.
\begin{table}[h]
\begin{center}
\caption{Averaged temperature values for the different CMOS sensors, during the characterization measurements.}
\begin{tabular}{lc}
\hline\\[-1.8ex]
   & T ($^o$C) \\
\hline\\[-1.8ex]
A  &   34.9$\pm$0.2 \\[0.25ex] 
B  &   33.1$\pm$0.2 \\ [0.25ex]
C  & 35.2$\pm$0.2 \\ [0.25ex]
D  & 40.1$\pm$0.5 \\ [0.25ex]
E  & 35.1$\pm$0.1 \\ [0.25ex]
F & 34.9$\pm$0.1 \\ [0.25ex]
\hline\\[-1.1ex]
\end{tabular}
\label{tab: measured_temp}
\end{center}
\end{table}

Sample D was always at a higher temperature of 40.1$\pm$0.5$^o$C.
As we do not have a measurement of its temperature when the sensor was still flat, we do not know if this characteristic was introduced during the curving process.  
This different operating temperature might be due to intrinsic properties of the die.
For sample B we have to consider that the measured temperature of 33.1$\pm$0.2$^o$C (in Table~\ref{tab: measured_temp}), corresponds to 35.1$^o$C, as its plastic substrate creates a $\sim2^o$C bias between the real surface temperature and the measured one (see Section~\ref{sec:cal_temp_res}).

\subsection{Methodology adopted for the characterization of curved CMOS}
\label{sec:met2}

As specified in Section~\ref{sec:data_acq} a set of 30 frames were acquired for each exposure time in the dark frames.
From these, a median image was obtained and a mean dark signal level was estimated from the average over all pixels. 
The dark current and the bias level were, hence, obtained by fitting these estimates as function of the exposure time.
 
The temporal noise, described in Section~\ref{sec:meas_quant}, was evaluated by building an image of the standard deviation of the 30 frames.
If in the previous case we computed the median image, this time we estimated the standard deviation, among the the 30 frames, of each pixel in the image, creating a single image made of standard deviations. 
Then, we obtained the temporal noise for a specific exposure time, by averaging over the pixels of this standard deviation image.
From this we obtained the RON as explained in Section~\ref{sec:meas_quant}.

We applied the same process to create the median image/mean signal level and the standard deviation image/temporal noise, also from the flat field frames.
The measured temporal noise and the mean signal level for several exposure time values were, thus, fitted according to the linear relation in Equation~\ref{eq:temp_noise}, and the gain, $k$, was obtained.

The noise of an image contains also the PRNU, which does not depend on time. 
Thus,by subtracting two frames exposed to the same uniform light level for the same amount of time and measuring the noise of this resulting image, we obtained the PRNU noise and the PNRU factor, as detailed in Section~\ref{sec:meas_quant}.
However by using only two frames we still have the influence of the temporal noise.
We suppressed this effect by subtracting a randomly extracted frame to each of the other 29 frames acquired per exposure time and then we computed the noise, $\sigma_{\mathrm{d}}$.
With the average of the 29 $\sigma_{\mathrm{d}}$, scaled by a factor of 2, we estimated the shot noise.
This, thus, led to $\sigma_{\mathrm{PRNU}}$ and $f_{\mathrm{PRNU}}$, which is usually expressed as a percentage of the mean signal (Equation~\ref{eq:tot_noise}).

\section{Results}
\label{sec:results}

In this Section we show the results of the characterization of all the curved samples and we compare them to the ones from the flat sample.
We use this comparison to statistically asses the impact of the curving process on the performances of the detectors.

\subsection{Measured dark current and RON}
\begin{figure*}[ht]
 \begin{center}
  \includegraphics[width=.98\textwidth]{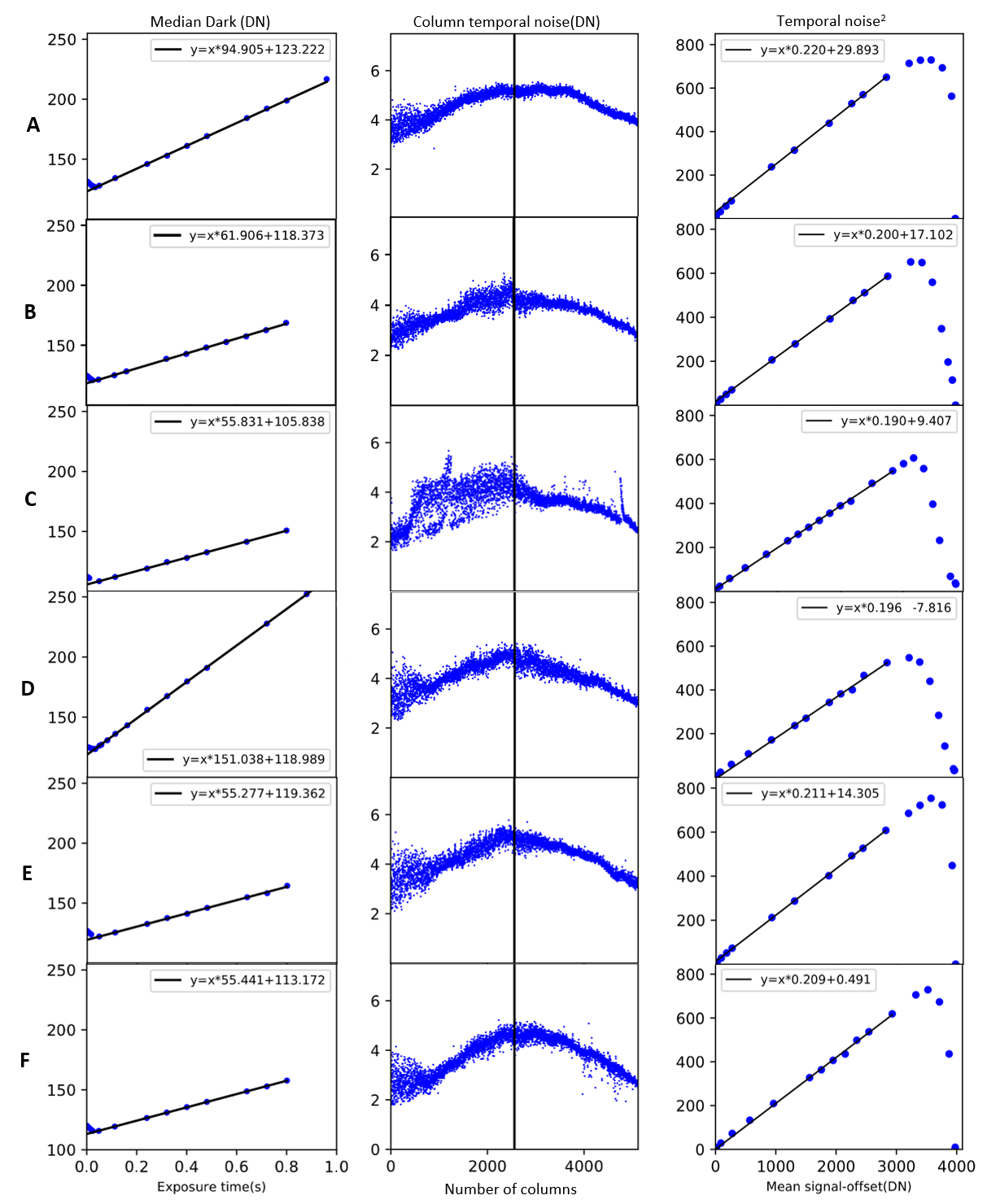}
  \caption{Left column: the blue solid circles are the median values of the dark exposure frames as function of exposure time, the black solid lines are the fits to the data. Middle column: column temporal noise vs column number of the sensor. The vertical black line indicates the column in the middle of the frame. Right column: 
  squared temporal noise of flat field frames as function of mean signal-offset (bias level). The black lines are the fits to the data for mean signal-offset between 1000\,DN and 3000\,DN. Each row of the Figure represents the measured quantities for a single detector (specified on the left side of the Figure).}
  \label{all_results}
 \end{center}
\end{figure*}

On the left column of Figure~\ref{all_results} are shown the dark current measurements (as described in Section~\ref{sec:met1}) and the linear fits for all samples (each row in the Figure~\ref{all_results} is a different detector).
For exposure times larger than 0.048\,s the measured signal increases linearly, as the charges due to the dark current accumulate in the pixels.
The black line in the plots are the linear fits to the data, from which the dark current value in DN/s (the slope of the fit) and the bias level (the intercept of the fit) are obtained.

After applying the gain to each detector (see Section~\ref{sec: measured_gain}), we find the dark current values in Table~\ref{tab: values}.
As already mentioned these measurements were performed at a temperature of $\sim35^o$C for all samples (except for sample D) and the dark current values for the curved detector samples are consistently smaller than the one of the flat detector.
More specifically they differ of: sample B 28\%, sample C 32\%, sample E 39\% and sample F 38\%.
The higher temperature of sample D makes the comparison of its dark current with the dark current of the other samples harder, but as most of the detector characteristics do not depend on temperature, we anyway show its results for completeness. 
The errors associated to the dark current in Table~\ref{tab: values} are the 1$\sigma$ errors on the linear fits.

The plots on the left column of Figure~\ref{all_results} show that, for very short exposure times, the counts on the sensors increase and the responses are not linear.
This feature is found in all sensors, thus, we concluded that it is an intrinsic characteristic of the CMV20000 CMOS.
This also implies that the measured value of the bias level is higher than the value from the fit of the dark current.
We used this higher value for the bias level in the rest of the paper (it is also the one written in Table~\ref{tab: values}), as we followed the definition of bias: the mean value of the median frame with the shortest exposure time acquired in darkness.

From the median image of the dark exposure with the shortest exposure time, we additionally evaluated the column temporal noise, by computing the standard deviation from the mean value of each column in the image.
These results are plotted vs column number in the middle column of Figure~\ref{all_results}.
The column temporal noise of the curved samples does not present any large variation with respect to the flat sample case, and it shows the same behavior with increasing noise values towards the center and decreasing values at the edges.
The measurements of the curved sensors also present mostly smaller values with larger scatter, compared to the flat sensor values.

The column temporal noise of the C sample shows significantly larger scatter due to the presence of one or more hot pixels per column.
The surface of this sample is slightly deformed, especially around the center.
The deviation from a perfect sphere is visible from Figure~\ref{four_prototypes}.
This, however, does not imply that the surface deformation has an impact on the noise of the detector, as the dark current did not show any sign of difference with respect to the other samples.
As the quality of the wafer of this specific chip was graded lower with respect to the other detectors presented here, those hot pixels could have been present even before the thinning and curving process.
We can not draw conclusive reason for this peculiar behavior, since the properties of the sensors, before the curving process, are unknown.

The RON was estimated from the temporal noise of the dark exposures acquired at the 
shortest exposure time, as explained in Section~\ref{sec:met1}.
These values are shown in Table~\ref{tab: values}.

\subsection{Measured gain, dynamic range and full well capacity}
\label{sec: measured_gain}
The gain was measured as explained in Section~\ref{sec:met1}, from a set of flat fields where the sensors are exposed to uniform and stable illumination.
The average signal measured on the frames grows linearly for exposure times larger than 0.048\,s until it reaches the saturation limit of 4095\,DN (set by the Analog Digital Converter, 12-bit per pixel) and plateaus. 
The saturation limit is specified in Table~\ref{tab: values} for all the detectors tested.
All of them reach it at 4095\,DN except for sample C that saturates at 3951\,DN.
By using the measured saturation limit, RON and bias level, we obtained the values (in Table~\ref{tab: values}) of dynamic range and full well capacity as in Equation~\ref{eq:fwc}.

The right column of Figure~\ref{all_results} shows the squared values of the temporal noise of the detectors against the mean signal subtracted by the bias level \citep{janesick_2007}.
The linear trend due to the accumulation of the charges inside the pixels (described by Equation~\ref{eq:temp_noise}), appears for values of mean signal - offset between 1000\,DN and 3000\,DN.
The black lines in the plots on the right column of Figure~\ref{all_results} are the fits to the data from which we obtain the gain values (the slope of the fits) in Table~\ref{tab: values}.
The errors here are the 1$\sigma$ errors on the linear fit.

The gain values of the curved samples differ from the one of the flat sample by 10\%, 16\%,  12\%, 5\% and 5\% respectively.
As the gain quoted by the manufacturer\footnote{\url{http://www.cmosis.com/products/product_detail/cmv20000}} is of 0.25 DN/e$^-$, 12\% larger than the gain of the flat sensor measured in this paper, the variation found among the gain values is considered within the manufacturing scatter and therefore not pointing to any specific effect due to the curving process.

\begin{table*}[h]
\begin{center}
\caption{Values for the electro-optical characterization of the flat and curved CMV20000 CMOS sensors. Note that the same methodology has been applied to all sensors and that sample D was measured at a different temperature with respect to the others.}
\begin{tabular}{lcccccc}
\hline\\[-1.8ex]
   & A  & B & C & D & E & F \\
\hline\\[-1.8ex]
Shape  & Flat & Concave & Concave & Convex & Convex & Concave \\[0.25ex] 
 $R_\mathrm{c}$ (mm)  &  $\infty$ & 150 & 150 & 280 & 280 & 170 \\[0.25ex] 
Bias (e$^-$)  &    595.9$\pm$24.2   & 622.8$\pm24.2$ & 588.4$\pm$21.7& 637.9$\pm$24.5 & 603.5$\pm24.8$ & 574.2$\pm$22.7 \\[0.25ex] 
Dark current  &   431.4$\pm$2.7    & 309.5$\pm$3.4 & 293.8$\pm$2.9 & 770.6$\pm$2.3 & 263.2$\pm$3.3 & 265.3$\pm$1.0 \\ [0.25ex]
(e$^-$/s) @ 35$^o$C & & & & @ 40$^o$C & & \\ [0.25ex]
Gain (DN/e$^-$)  & 0.220$\pm$0.003 & 0.200$\pm$0.002 & 0.190$\pm$0.002 & 0.196$\pm$0.006 & 0.210$\pm$0.002 & 0.209$\pm$0.005 \\ [0.25ex]
RON (e$^-$)  & 11  & 10 & 10 & 10 & 10 & 10 \\ [0.25ex]
Saturation (DN)  & 4095 & 4095 & 3951 & 4095 & 4095 & 4095 \\ [0.25ex]
Dynamic range (dB) & 64.74 & 66.26 & 66.44 & 66.26 & 65.98 & 66.14 \\ [0.25ex]
Full well (e$^-$) & 18018 & 19852 & 20206 & 19331 & 18896 & 19019\\ [0.25ex]
PRNU factor & 1.2\% & 2.0\% & 2.1\% & 1.9\% & 2.0\% & 1.9\%\\ [0.25ex]
\hline\\[-1.1ex]
\end{tabular}
\label{tab: values}
\end{center}
\end{table*}

The last characteristic quantity shown in Table~\ref{tab: values} is the PRNU factor, $f_{\mathrm{PRNU}}$, computed as explained in Section~\ref{sec:met2}.
The PRNU factor values do not present large differences among the samples, flat and curved.
\section{Conclusions}
\label{sec:conclusion}
The recent progress made towards the manufacturing of curved detectors represents a step forward to the creation of more compact and performing optical systems with very wide applications.
It also opens the possibility to new designs, whose realization was technologically impossible before.
Here, we presented the characterization results of a sample of five curved CMOS detectors developed within a collaboration between CEA-LETI and CNRS-LAM.
These detectors are CMV20000 with 20\,Mpix (manufactured by CMOSIS) and they have been curved with different shapes and radii of curvature over the full sensitive area of 24x32\,mm$^2$.
The curved detector sample is composed of two concave curved down to $R_\mathrm{c}=$150\,mm, another concave with $R_\mathrm{c}=$170\,mm and two convex with $R_\mathrm{c}=$280\,mm.

Since the packaging of the detectors is the same as the original one, it is possible to plug the detectors in directly in any interface or camera that was already built for the flat off-the-shelf sensor.
In order to perform the tests and readout the detectors, we used the CMV20000 Evaluation Kit (from CMOSIS).
This also allowed us to readout the internal temperature sensors of the CMV20000 chips.

These temperature sensors were calibrated against four thermocouples for all the detectors tested in this paper (Section~\ref{sec:temp_cal_intro}).
The thermocouples were glued at the back of small copper blocks, that were in turn thermally linked to the back of the ceramic support located below the detector sensitive area.
Such thermocouples have been readout at the same time of the internal temperature sensor.
The measured temperature was sampled from room temperature ($\sim19^o$C) until full thermalization of the sensors ($\sim35-40^o$C).
Once the calibration of the internal temperature sensors was achieved, the thermocouples were removed.

In Section~\ref{sec:met1} we described the quantities to characterize, the setup used, and the methodology for performing the measurement and analyzing the results.
The same characterization steps were repeated for all the sensors in the test sample.
From Table~\ref{tab: values} we have an overview of the results and we find them to be mostly homogeneous between the flat and curved samples.
A large difference (with respect to the other detectors) in dark current is measured for sample D (one of the convex shaped).
For this sample the dark current is almost 3 times larger than the dark current measured for the others, and this is due to the larger temperature that the sensor reached while performing its characterization.
While all the other samples reached a stable characterization temperature of $\sim35.0^o$C, sample D was tested at $\sim40.0^o$C.

When we compared the dark current value for the flat sensor, to the dark current measured for the other samples (excluding sample D), we obtained substantially lower values for the latter ones: from 28\%, up to a maximum of 39\% difference.
A similar decrease of dark current in curved CMOS sensors was already found in other works \citep{Itonaga_2014,guenter2017} and has been attributed to an alteration of the band gap in the sensitive area of the devices due to the strain induced in the curving process. As in our case the curving process is made without fixing the edges (as in \cite{guenter2017}) and the dimension of the sensors is much larger than their thicknesses, some decrease in dark current is expected in our spherically curved samples \citep{gregory_2015}.
However it is not excluded that some amount of this difference might be due to intrinsic properties of the samples themselves. This could also explain their similar dark current values.

We also measured a smaller readout noise of 10\,e$^-$ for all the curved sensors with respect to the 11\,e$^-$ for the flat sensor.
This smaller RON generates a larger dynamic range $>$66\,dB, against the 64.74\,dB of the flat sensor.
We find no significant difference in the bias level, as the values mostly match within the errors.
We also find similar behavior of the column temporal noise between all sensors, where the curved samples presented smaller values, with larger scatter, compared to the flat one.
From the measurements, the gains show a discrepancy from 5\% to 16\% between the curved sensors compared to the flat one, which might be due to an intrinsic characteristic that the chips already had before curving them. 

The PRNU factors of the curved samples show an increase of $\sim$0.8\% with respect to the value for the flat sensor.
The difference between these is not significant.
This also holds true for sample C, that even having a deformed shape, does not present any particular degradation in its performances. 

From the overall performances tested in this paper, we conclude that the curving process shown here does not impact the main electrical characteristics of the detectors and in some cases, e.g. the dark current, it might even improve them.
The astronomical community can particularly gain from this characteristic. 

The next step for this work is to verify the quality of the detector surface, and if necessary, improve the curving technique until it reaches the required precision for astronomical applications.
Once this is achieved, a new prototyping phase can begin to develop curved CCDs.

Funding Information. ERC (European Research Council)
(H2020-ERC-STG-2015-678777
ICARUS program), the French Research Agency (ANR) through the LabEx FOCUS ANR-11-LABX-0013.
Acknowledgments. The authors acknowledge the support
of the European Research council through the H2020-ERC-STG-2015-678777
ICARUS program.


\bibliography{sample}

\bibliographyfullrefs{sample}


\end{document}